\documentclass[reprint,amsmath,amssymb,aps]{revtex4-1}

\usepackage{graphicx} 
\usepackage{dcolumn} 
\usepackage{bm}
\usepackage{natbib}

\begin{document}

\preprint{APS/123-QED}

\title{Squeezed-twin-beam generation in strongly absorbing media}
 
\author{Jon D. Swaim}
\author{Ryan T. Glasser}
 \email{rglasser@tulane.edu}
\affiliation{Physics and Engineering Physics Department, Tulane University, New Orleans, 70118, USA}
 
\date{\today}

\begin{abstract}
We experimentally demonstrate the generation of squeezed, bright twin beams which arise due to competing gain and absorption, in a medium that is overall transparent.  To accomplish this, we make use of a non-degenerate four-wave mixing process in warm potassium vapor, such that one of the twin beams experiences strong absorption.  At room temperature and above, due to Doppler broadening and smaller frequency detunings compared to other schemes, the ground state hyperfine splittings used in the present double-$\Lambda$ setup are completely overlapped.  We show that despite the resulting significant asymmetric absorption of the twin beams, quantum correlations may still be generated.  Our results in this new regime demonstrate that the simplified model of gain, followed by loss, is insufficient to describe the amount of quantum correlation resulting from the process.  
\end{abstract}

\pacs{Valid PACS appear here} 
                             
%\keywords{squeezed light, quantum optics, potassium, atomic vapor}

\maketitle

%%%%%%%%%%%%%%%%%%%%%%%%%%%%%%%%%%%
% Introduction and basic physics
%%%%%%%%%%%%%%%%%%%%%%%%%%%%%%%%%%%
\section{Introduction}

Quantum theory has led to the prediction and experimental realization of several novel states of light which are unobtainable by classical theory alone~\cite{Glauber1963, Walls1983, Walls2008}.  Of particular interest are squeezed states~\cite{Walls1983, Kumar1984,Peng_2000}, in which the noise in one quadrature of the electromagnetic field can be reduced beyond the quantum limit, i.e., shot noise, at the expense of introducing additional noise in the orthogonal quadrature.  The first demonstration of such a quantum effect was by Slusher et al., with squeezed light generated via non-degenerate four-wave mixing (4WM) with Na atoms in an optical cavity~\cite{Slusher1985}.  Squeezed light has since been demonstrated with optical parametric oscillators~\cite{Camy_1987,Wu1987, Furst2011}, optical fibers~\cite{Levenson1985, Leuchs_fiber} and mechanical resonators~\cite{Safavi-Naeini2013}.  Two-mode squeezed states are a particular class of squeezed light in which quantum correlations are generated in the difference or sum of quadratures between the twin beams, often produced via parametric down conversion in crystals~\cite{Wu1986} or 4WM in atomic vapor~\cite{McCormick2007, Guo2014,Vogl_NJP,Peng_1992}.  In these systems, strong reductions in noise between joint-quadratures have enabled demonstrations of continuous variable entanglement~\cite{Reid1988, Kwiat1995, Boyer2008}.  In addition, atomic vapors have been labeled as excellent candidates for high-dimensional entanglement, due to their inherent multi-mode nature and large gain~\cite{Boyer2008}.  When the aforementioned systems are seeded with a coherent input probe, the processes generate twin beams that are quantum correlated in their relative intensities, resulting in intensity-difference squeezing.

\begin{figure}[b!]
\begin{center}
\includegraphics*[width=\columnwidth]{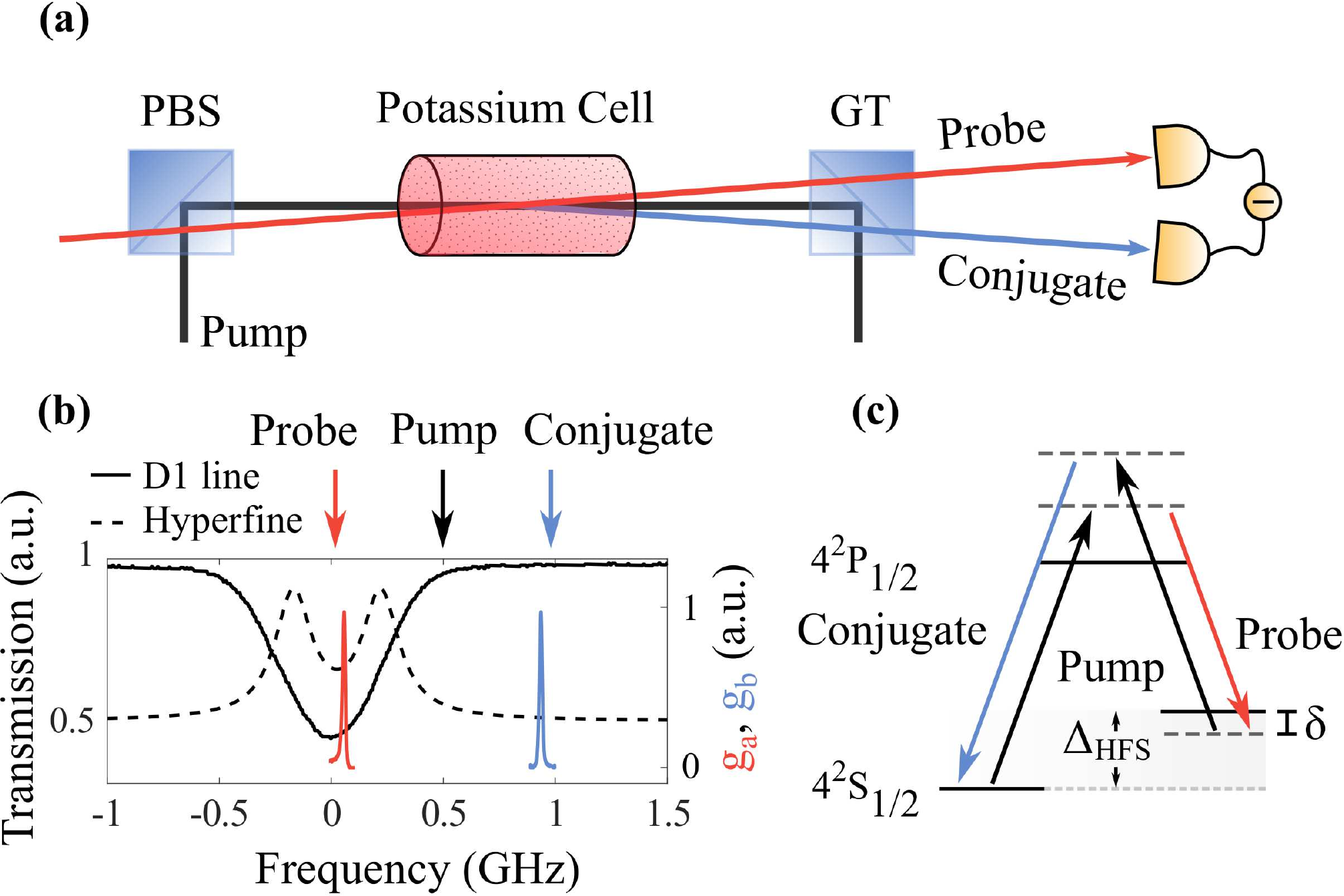}%{FIG1v2.eps}
\caption{\label{setup}Four-wave mixing in potassium vapor.  (a) The pump and probe fields overlap in the center of the vapor cell, and the amplified probe and generated conjugate fields are detected on a balanced photodetector.  PBS: polarizing beam splitter.  GT: Glan-Taylor polarizer.  (b)  Measured absorption spectrum of the $^{39}$K D1 line (solid line), a double-Lorentzian model of the hyperfine states based on measurements of the transitions using polarization spectroscopy (dashed), and the measured gain lines for the probe (red) and conjugate (blue).  The arrows indicate the frequency detunings used in the experiment.  (c) Energy level diagram summarizing the four-wave interaction.}
\end{center} 
\end{figure}

It is well known that quantum correlations, including those discussed above, are susceptible to the adverse effects of decoherence through interactions with the environment, often in the form of absorption or loss~\cite{Zurek_review, Kiss1995}.  For this reason, approaches based on atomic vapors have found much success using off-resonant schemes such as the double-$\Lambda$ configuration~\cite{McCormick2007}.  This setup achieves strong coherences between the atomic levels involved, and results in negligible absorption of the twin beams.  McCormick et al. first proposed this method, and demonstrated 3.5 dB of relative intensity squeezing between twin beams generated via 4WM in hot rubidium vapor~\cite{McCormick2007}.  In subsequent experiments, the method produced as much as 9.2 dB of squeezing~\cite{McCormick2008, Glorieux2010a}.  We note that these 4WM schemes often allow for a simpler implementation than parametric down conversion in nonlinear crystals, since large values of squeezing can be obtained without the use of a cavity, periodic poling, or waveguide techniques.  Similar results have been obtained using cesium atoms~\cite{Guo2014}.  These two systems share the property that the ground state hyperfine splitting is resolved at operating temperatures.  More recently, 4WM in potassium vapor was investigated on account of its enhanced $\chi^{(3)}$ susceptibility and large intrinsic gain~\cite{Zlatkovic2016}, though squeezed light was not demonstrated.  Unlike rubidium and cesium, potassium's hyperfine splitting $\Delta_{\textrm{HFS}} = 462$ MHz is on the order of the Doppler-broadened linewidth, and thus all of the atomic transitions are completely overlapped at room temperature. Consequently, the probe field is almost maximally absorbed.  While the effect of loss on squeezed light has been studied in the context of off-resonant twin beam generation~\cite{McCormick2008, Glorieux2010, Jasperse2011}, it remains to be elucidated whether this strong asymmetric absorption would preclude the ability to generate quantum correlation. 

We demonstrate for the first time a squeezed light source based on potassium vapor, in which one of the twin beams is positioned at the center of a Doppler-broadened absorption profile, where the absorption is highly asymmetric.  To achieve this, we implement a double-$\Lambda$ scheme for the D1 line of $^{39}$K, and show a 1.1 dB reduction in relative intensity noise below the shot noise limit (SNL).  Following the approach developed in Ref.~\cite{Jasperse2011}, we show that such strong quantum correlations are not expected when the absorption is modeled solely as loss associated with a beam splitter interaction that follows an ideal gain medium.  Rather, it is the competition between the gain and absorption within the medium which leads to accurate predictions of squeezing.  This important difference permits the survival of quantum correlation when appropriate operating parameters are chosen, even in the presence of significant asymmetric loss between the twin beams.  

%%%%%%%%%%%%%%%%%%%%%%%%%%%%%%%%%%%
% Experimental setup
%%%%%%%%%%%%%%%%%%%%%%%%%%%%%%%%%%%
\section{Experimental setup}

The essence of the experimental setup is shown in Fig.~\ref{setup}a.   The 4WM process is pumped with 380 mW of coherent light from a Ti:Sapph continuous-wave laser, detuned approximately $500$ MHz to the blue side of the $^{39}$K D1 line (Figs.~\ref{setup}b-c).  The pump is collimated to a size of $0.8$ mm $\times$ $0.6$ mm and then passed through an 80 mm, anti-reflection-coated vapor cell containing natural-abundance potassium.  The vapor cell is heated to approximately $100$ $^\circ$C, and the pump is filtered out after the cell using a Glan-Taylor polarizer.  The probe is generated by sending a fraction of the light through an acousto-optic modulator, double-passed and driven at a frequency of $\Omega/2\pi = 238$ MHz.  The orthogonally polarized pump and probe ($1/e^2$ diameter $\approx670$ $\mu$m) beams are then combined on a polarizing beam splitter (see Fig.~\ref{setup}a), and overlap in the center of the cell at a small angle ($\theta \sim 0.1^\circ$).  We note here that this angle is smaller than in Rb and Cs-based 4WM schemes \cite{Turnbull2013, Zlatkovic2016}.  As a result of the 4WM process, a conjugate beam is generated at an angle $2\theta$ relative to the probe, with frequency $\omega_0 + \Delta_{\textrm{HFS}}+\delta$ (Figs.~\ref{setup}b-c), where $\omega_0$ is the pump frequency and the single photon detuning $\delta$ is controlled by varying $\Omega$.  In Fig.~\ref{setup}b, we also show the measured gain lines for the probe and conjugate, and indicate the relevant frequency detunings of all beams relative to the hyperfine states, as well as the Doppler-broadened absorption profile.  Both the probe and conjugate beams are spatially filtered using irises to remove any residual pump light, and redirected to a balanced photodetector (Thorlabs PDB450) via two D-mirrors and two lenses.  Lastly, the signals are amplified with a gain of 10$^5$ V/A and the difference signal is analyzed on a spectrum analyzer using a resolution bandwidth of 10 kHz and a video bandwidth of 30 Hz.

%%%%%%%%%%%%%%%%%%%%%%%%%%%%%%%%%%%
% Main results
%%%%%%%%%%%%%%%%%%%%%%%%%%%%%%%%%%%
\section{Results}

\begin{figure}[b!]
\begin{center}
\includegraphics*[width=\columnwidth]{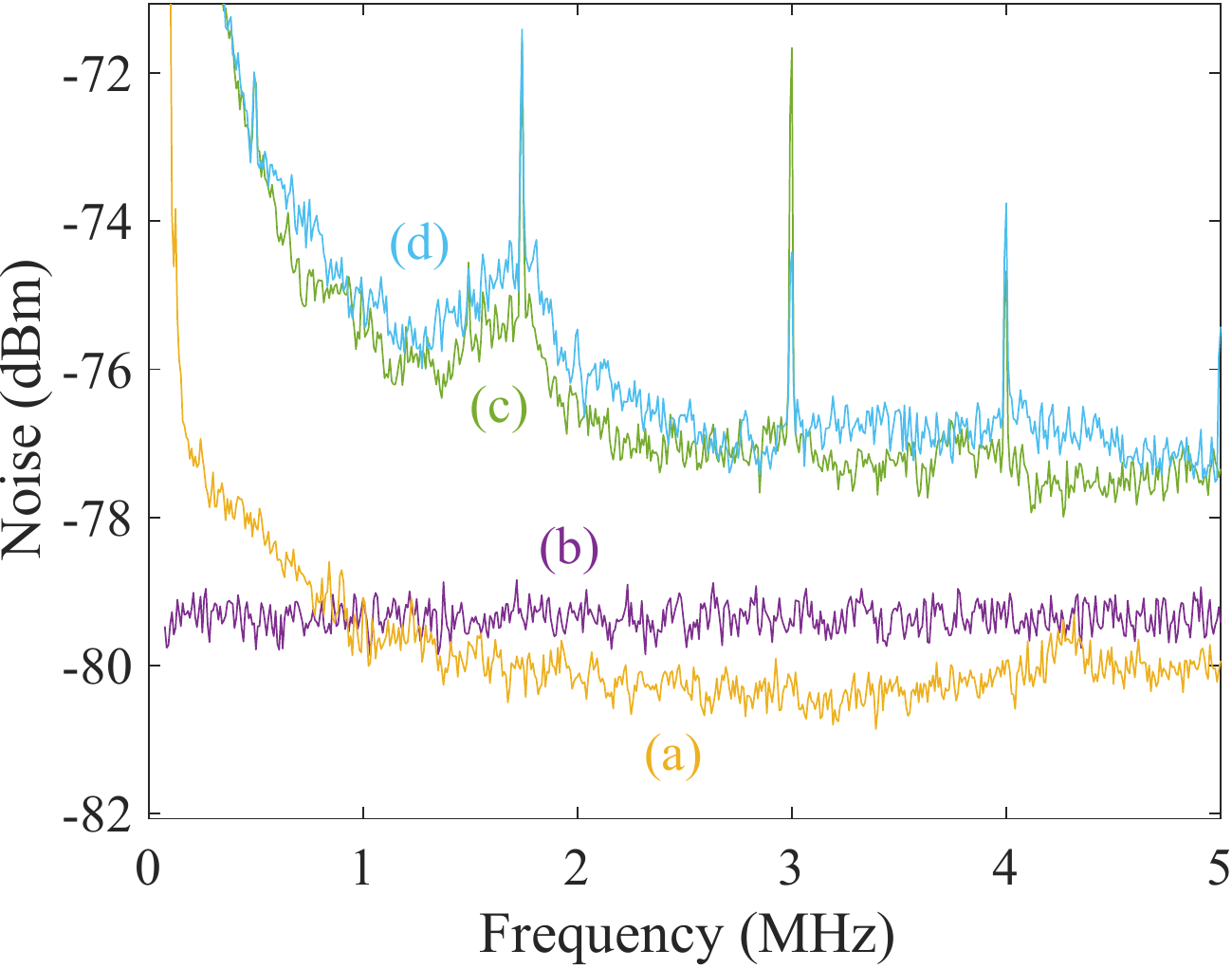}%{FIG2v2.eps}
\caption{\label{SA_results}Two-mode squeezing via 4WM in potassium vapor.  The noise power spectra correspond to (a) the 4WM difference signal for 74 $\mu$W of probe and 62 $\mu$W of conjugate beam powers, (b) the shot-noise limit for 136 $\mu$W of pump light incident on a 50/50 beamsplitter and the individual spectra for (c) 62 $\mu$W of conjugate and (d) 74 $\mu$W of probe, respectively.}
\end{center}
\end{figure}

\begin{figure}[t!]
\begin{center}
\includegraphics*[width=\columnwidth]{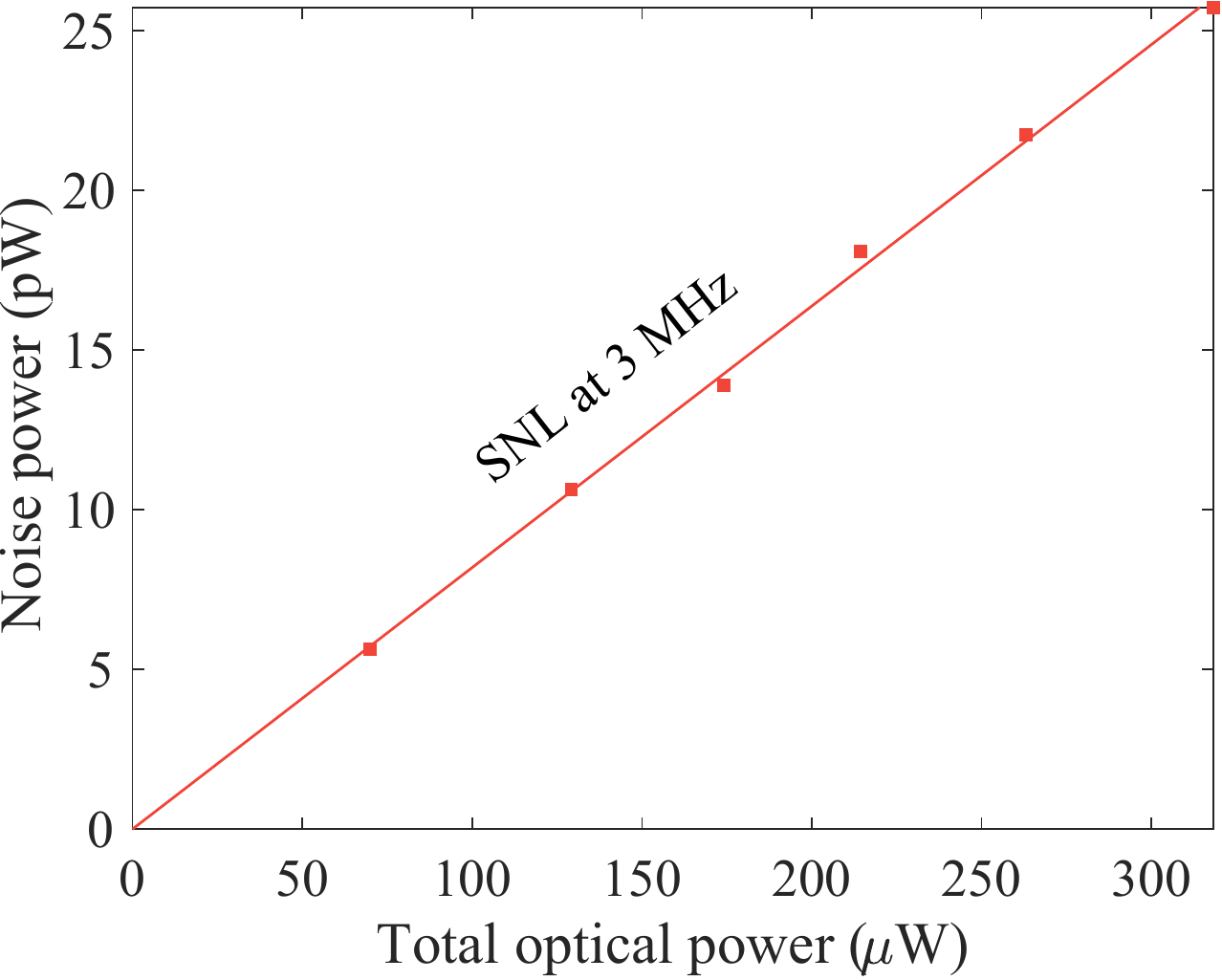}%{SFIG1.eps}
\caption{\label{shotnoise} Measured noise power of our laser light at 3 MHz as a function of the total optical power.  The data has been corrected for an offset due to electronic noise.}
\end{center} 
\end{figure}

One of the main results is shown in Fig.~\ref{SA_results}.  The noise power spectrum labeled (a) displays the 4WM difference signal measured for $P_a = 74$ $\mu$W of probe and $P_b = 62$ $\mu$W of conjugate beam powers.  For this result, we have used an input power of $P_{\textrm{in}} = 120$ $\mu$W, and estimate the effective probe and conjugate gains within the medium (accounting for iris attentuation and other optical losses with $\eta^\prime_{a,b}$) to be $g_a = P_a/ (\eta^\prime_a P_{\textrm{in}}) = 1.9$ and $g_b = P_b/ (\eta^\prime_b P_{\textrm{in}}) = 0.7$, respectively.  The difference signal falls below the SNL (b) for frequencies between 900\,kHz and 5\,MHz (roughly the bandwidth of the detector), and reaches a minimum of 1.1 dB below the SNL at $3$ MHz.  We confirmed that the laser light in (b) was shot-noise limited by measuring the noise power as a function of total optical power (see Fig.~\ref{shotnoise}).  Also, the noise powers of the individual conjugate and probe beams, labeled (c) and (d) in Fig.~\ref{SA_results}, are more than 5 dB above the SNL for a single beam (i.e., 3 dB below the SNL trace shown in Fig.~\ref{SA_results}).  Finally, as a check, we have confirmed that both the residual pump noise (obtained by blocking the input probe) and electronic dark noise are below the 4WM signal and the SNL, with a minimum clearance of 4 dB.

%We also show the SNL for an equivalent amount of total optical power, denoted by (b), which is obtained by evenly splitting 136 $\mu$W of pump light and redirecting it to the balanced photodetector using flip-mirrors.  

We found that the best squeezing could be obtained with the pump detuned approximately 500 MHz to the blue side of the broadened D1 line (solid curve in Fig.~\ref{setup}b), and with $\delta \sim 15$ MHz.  In terms of the double-$\Lambda$ scheme, this situation is similar to that used previously with rubidium~\cite{McCormick2007} and cesium~\cite{Guo2014}, with the probe frequency lying in between the hyperfine ground states.  However, in the present case, the probe is positioned so that it is centered on the Dopper-broadened profile and almost maximally absorbed.  The probe transmission $t$ is calculated as the ratio of the output and input powers of the probe with the pump blocked, taking into account optical losses after the cell.  For the data shown in Fig.~\ref{SA_results}, the probe transmission is $t = 17\%$.  Additionally, since the measured value of $t$ near resonance depends on the choice of cell temperature and two photon detuning, we were able to measure squeezing with transmissions ranging from $12\%$ to $42\%$.  We observe that, in the case of rubidium or cesium, this configuration could involve a resonant pump and thus additional noise which would most likely preclude the observation of any degree of squeezing~\cite{McCormick2008, Jasperse2011}.

%%%%%%%%%%%%%%%%%%%%%%%%%%%%%%%%%%%
% Discussion 
%%%%%%%%%%%%%%%%%%%%%%%%%%%%%%%%%%%

\begin{figure}[b!]
\begin{center}
\includegraphics*[width=\columnwidth]{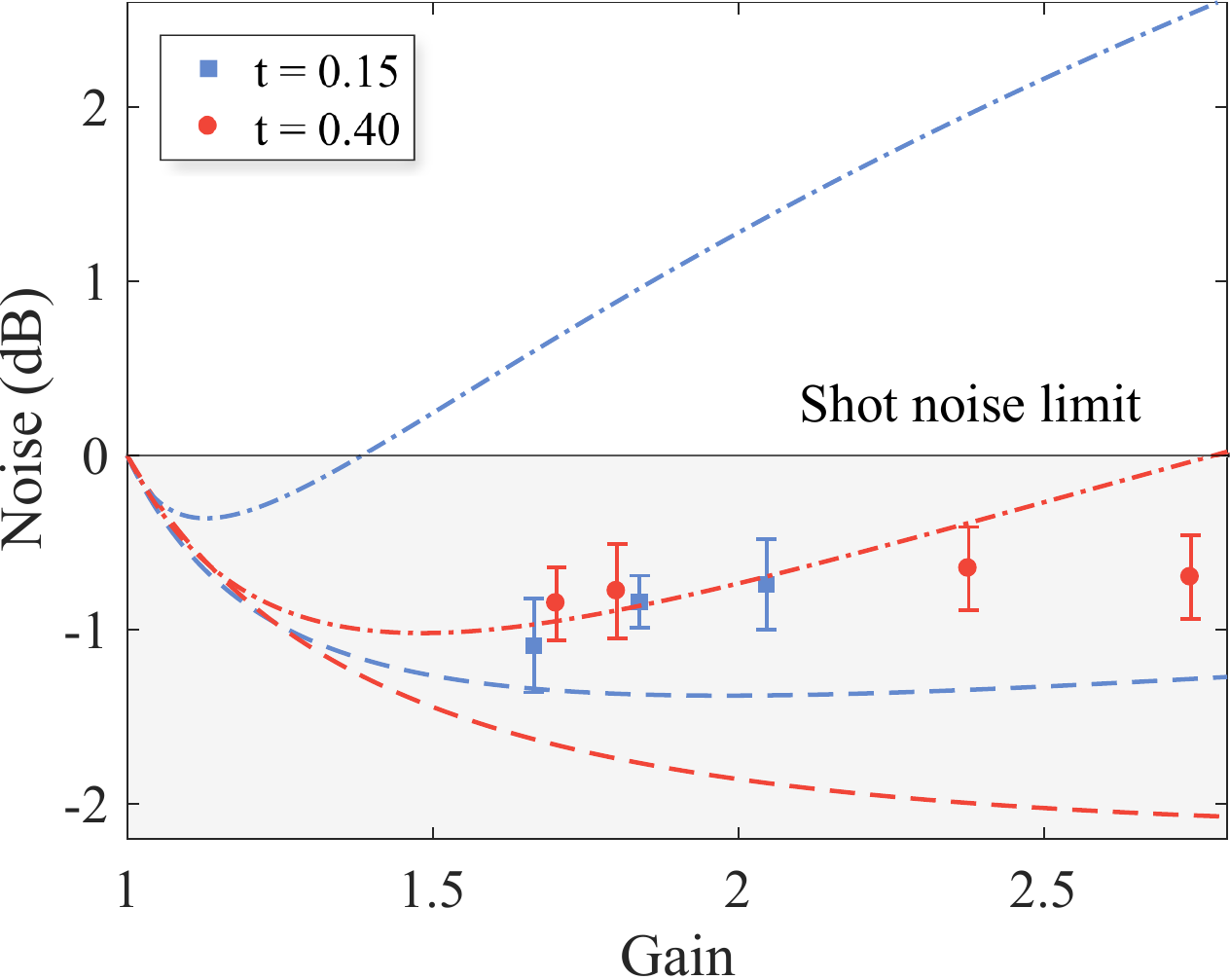}%{FIG3edit.eps}
\caption{\label{gaindata}Measured squeezing compared to the two theoretical models described in the text (DGL, dashed lines, and BS, dashed-dotted lines).  Data is shown for various levels of gain and measured probe transmissions of $t = 0.15$ (blue squares) and $t = 0.4$ (red circles), where each data point corresponds to the measured noise power at 3 MHz relative to the SNL.  Also shown are the predicted values of squeezing for transmissions of $15\%$ (blue) and $40\%$ (red) using the two theoretical models.}
\end{center}
\end{figure}

\begin{figure*}
\begin{center}
\includegraphics*[width=\textwidth]{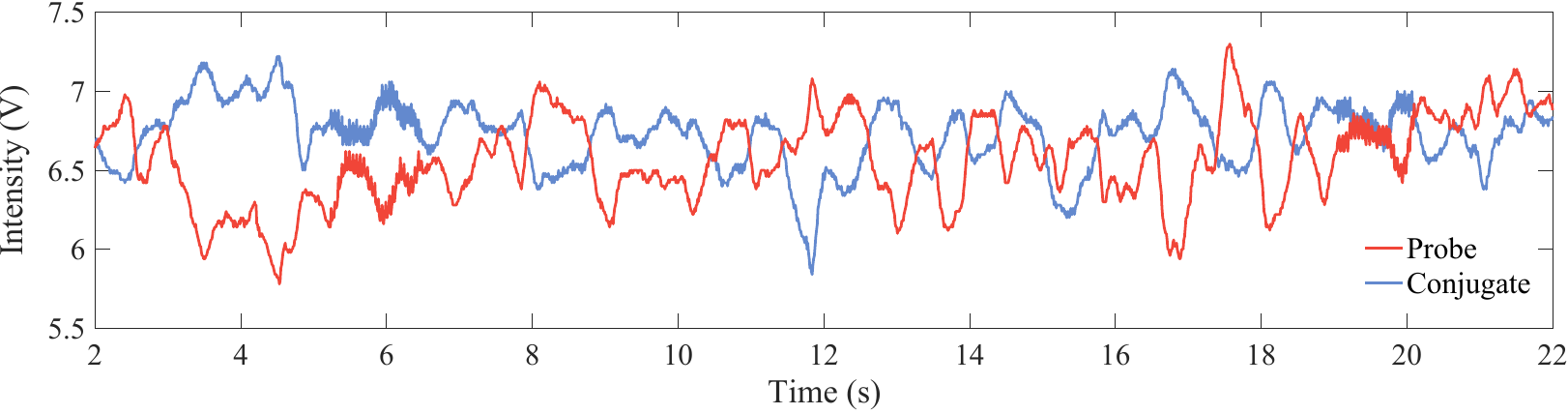}%{SFIG2edit.eps}
\caption{\label{anticorrelation} Anti-correlated fluctuations in the probe and conjugate intensities measured over a time scale of twenty seconds.  In this mesurement, the experimental parameters are the same as those which lead to the squeezing data shown in Fig.~\ref{SA_results}.}
\end{center} 
\end{figure*}

We now turn to investigating the ramifications of the strong absorption of the probe beam, as this is an important difference between the present scheme and previous double-$\Lambda$ experiments.  Intuitively, one might expect this asymmetric absorption to severely reduce, if not completely destroy, quantum correlations in the two-mode state.  This is often understood through means of a simple beamsplitter interaction, in which a fraction of the light is lost through one port to the environment and vacuum is injected into the other port~\cite{Kiss1995}.  In this model, beamsplitters are placed in both the probe and conjugate paths, and losses are introduced via detection efficiencies $\eta_a$ and $\eta_b$, respectively.  In the present case where absorption only occurs on the probe beam, the probe transmission is given by $t = \eta_a / \eta_b$ and the predicted squeezing in the beamsplitter (BS) model is
\begin{equation}
S_\textrm{BS} =  1+ \eta_b\frac{ 2(g-1)\left(g(t-1)^2-1\right)}{g(t+1)-1}
\label{eq:SBS}
\end{equation}
\noindent where $g$ is the intrinsic 4WM gain.  In this model, the squeezing scales much less favorably for asymmetric losses (i.e., $t \neq 1$), than if the losses were symmetric (or balanced)~\footnote{This is not strictly true in the case when the probe is seeded with a coherent state and the conjugate is seeded with vacuum.  In this case, a small amount ($\sim$1-5$\%$) of probe loss is actually optimal.}.  However, as the loss occurs $\emph{within}$ the medium itself, a competition between gain and absorption occurs, resulting in a different scaling of the squeezing parameter, as shown in the theoretical and experimental work of Refs.~\cite{McCormick2008, Jasperse2011}.  In this model of distributed gain and loss (DGL), the expression for squeezing given a probe transmission $t$ and balanced detection efficiency $\eta$ is ultimately more complicated than Eq.~\ref{eq:SBS}, but can be summarily expressed as
\begin{equation}
S_{\textrm{DGL}} =  1 - \eta \tilde{S}_\textrm{4WM} + \eta \sqrt{t} \tilde{S}_\text{vac}
\label{eq:SDGL}
\end{equation}
\noindent where $\tilde{S}_{\textrm{4WM}}$ accounts for correlations which reduce the noise below the SNL, and $\tilde{S}_{\textrm{vac}}$ introduces additional vacuum noise.  The analytical expressions for $\tilde{S}_{\textrm{4WM}}$ and $\tilde{S}_{\textrm{vac}}$ are given in Ref.~\cite{Jasperse2011}, and only depend on $g$ and $t$ when the conjugate absorption by the atoms is negligible.    

In Fig.~\ref{gaindata}, we show the measured squeezing at various levels of intrinsic gain $g = g_b + 1$ for two different regimes of probe transmission: $t = 0.15$ (blue squares) and $t = 0.4$ (red circles).  We have also included calculations of the predicted squeezing for these two regimes, using both the BS model of Eq.~\ref{eq:SBS} (dashed-dotted) and the DGL model of Eq.~\ref{eq:SDGL} (dashed).  In the experiment, the combination of optical and detection losses leads to an overall detection efficiency of $\eta = 0.5$.  Because of this, we take $\eta_b = 0.5$ and $\eta = 0.5$ in the two models, BS and DGL, respectively.  Importantly, we see that squeezing is not expected for gains $\geq 1.4$ when the probe transmission is $t = 0.15$ in the BS model (blue dash-dotted curve).  Rather, due to the asymmetric losses, the BS model predicts excess noise in this regime.  On the other hand, there is good qualitative agreement between the experiment and the DGL model in this case, with both showing quantum correlation between the resultant beams.  This agreement is noteworthy given that effects other than those considered in the model are expected to become important due to Doppler broadening.  Also, it should be pointed out that the irises used in mode selection~\cite{Boyer2008} attenuate some of the light, likely resulting in additional discrepancy.  

\section{Discussion and Outlook}

The Heisenberg uncertainty principle does not place a constraint on the maximum attainable value of squeezing.  In the presence of balanced detection loss ($1-\eta$), however, the squeezing is limited to $10\times \textrm{log}_{10} (1-\eta)$.  With unbalanced losses, the maximum achievable squeezing occurs for a particular, optimal value of intrinsic gain.  This is evident in the minima of the theoretical curves shown in Fig.~\ref{gaindata}.  In the DGL model, the optimal gains for our two experimental scenarios are $2$ and $3.8$, respectively.  The dependence of the optimal gain on the probe transmission offers an explanation as to why, in Fig.~\ref{gaindata}, the probe transmission differs by more than a factor of two and yet there is not an appreciable change in squeezing.  In general, the optimal gain increases with increasing probe transmission.  Experimentally, however, the available gain is limited by the maximum available pump intensity, as well as excess noise introduced at higher vapor cell temperatures.  Under completely ideal circumstances (i.e., optimal gain and $\eta = 1$), our calculations suggest that the attainable squeezing would be $3.4$ dB and $6.4$ dB for probe transmissions of $15\%$ and $40\%$, respectively.

An interesting corollary which is inherent to this system is that quantum correlation can be present even when the medium is effectively transparent.  This was demonstrated in Refs.~\cite{Jasperse2011, Glorieux2011}, with the loss occuring within the medium itself due to absorption in the former experiment, and by introducing loss after the 4WM process in the latter.  In this sense, the overall gain is equal to one, as the total input probe power equals the sum of the probe and conjugate output powers after the added loss.  In our experiment, we have observed $-0.84 \pm 0.16$ dB of squeezing when the normalized powers of the probe and conjugate beams are $0.56 \pm 0.05$ and $0.47 \pm 0.04$, respectively, giving an overall gain of $1.03 \pm 0.06$ (all errors are one standard deviation throughout this manuscript).  As we have shown that the beamsplitter-only loss model is insufficient in predicting the observable squeezing values when the 4WM medium exhibits strong absorption on one of the beams, in this regime we believe that competing gain/loss plays an important role in generating squeezed beams in an overall transparent medium.

Lastly, due to dispersive effects within the medium, a time delay exists between the output probe and conjugate beams which depends on various parameters of the 4WM process~\cite{Boyer2007, Glasser2012}.  This delay gives rise to a frequency dependence in the squeezing spectrum, typically limiting the maximum frequency at which squeezing is observed.  In our experiments, this frequency is above the bandwidth of the detector.  Additionally, we observe a somewhat related effect, in which the nature of the correlation is also found to be frequency dependent.  While fluctuations in the 4WM signal are correlated at high frequencies ($\geq$ 900 kHz) and lead to squeezing, we find that the intensity fluctuations at low frequencies ($\leq$ 1 Hz) are anti-correlated (see Fig.~\ref{anticorrelation}).  We suspect that the dispersive nature of the probe leads to a frequency dependence of the squeezing angle, in a manner similar to that described in Refs.~\cite{Kimble2001, Mikhailov2006, Corzo2013}, resulting in a rotation of the squeezing ellipse with frequency.  In future work, phase-sensitive measurements of the probe and conjugate modes could help to clarify this phenomenon, as well as characterize all joint-quadrature information in the present bright twin-beam case.   

%%%%%%%%%%%%%%%%%%%%%%%%%%%%%%%%%%%
% Summary and Outlook
%%%%%%%%%%%%%%%%%%%%%%%%%%%%%%%%%%%
\section{Conclusion}

In this work, we have experimentally investigated the behavior of quantum-correlated twin beams in a new regime, that in which strong absorption plays a significant role.  Our results indicate that a competition between gain and absorption exists within the medium, acting to reduce the deleterious nature of loss and thus improve quantum correlations.  Despite a measured probe transmission of only $17\%$ and a detection efficiency of $\eta = 0.5$, we have measured a reduction in noise of $1.10 \pm 0.27$ dB below the SNL.  Intrinsically, this configuration leads to such strong attenuation of the probe beam that we are able to demonstrate quantum correlations even when the medium has become effectively transparent, rendering the medium a quantum beamsplitter for photons as discussed in Ref.~\cite{Glorieux2011}.  In addition to broadening our understanding of dissipative quantum optics, these results may be relevant for the engineering of quantum-correlated states in other configurations with strong absorption.  Also, as this is the first demonstation of squeezing using potassium vapor, the results open up the possibility of interfacing squeezed light with cold atom experiments involving potassium, although significant effort to lower the frequency of obtained squeezing would likely be necessary, in order to achieve reasonable interaction times.  Finally, they could also provide enhancements in precision measurement when detection is performed near the potassium D1 line.

\section*{Acknowledgements}
The authors would like to acknowledge funding from the Louisiana State Board of Regents and Northrop Grumman $\emph{NG - NEXT}$.  We also thank Roger C. Brown for initial discussions surrounding the experiments performed here.

%\bibliographystyle{apsrev4-1}
%\bibliography{refs}

\end{document}